\documentclass[preprint]{vgtc}               

\ifpdf
  \pdfoutput=1\relax                   
  \usepackage{graphicx}                
  \DeclareGraphicsExtensions{.pdf,.png,.jpg,.jpeg} 
\else
  \usepackage{graphicx}                
  \DeclareGraphicsExtensions{.eps}     
\fi%

\graphicspath{{figures/}{pictures/}{images/}{./}} 

\usepackage{microtype}                 
\PassOptionsToPackage{warn}{textcomp}  
\usepackage{textcomp}                  
\usepackage{mathptmx}                  
\usepackage{times}                     
\usepackage{cite}                      
\usepackage{tabu}                      
\usepackage{booktabs}                  
\usepackage{balance}
\usepackage[table, dvipsnames]{xcolor} 

\usepackage{tabularx}
\usepackage{float}
\usepackage{subfigure}

\usepackage{todonotes}
\usepackage{balance}       
\newcommand{\ra}[1]{\renewcommand{\arraystretch}{#1}}
\usepackage{enumitem}

\usepackage{soul}

\usepackage[most]{tcolorbox}
\newtcolorbox{highlighted}{colback=yellow,breakable}

\onlineid{8820}

\vgtccategory{Research}

\vgtcinsertpkg

\title{How to evaluate data visualizations across different levels of understanding}

\author{Alyxander Burns\thanks{e-mail: alyxanderbur@cs.umass.edu}\\ %
        \scriptsize UMass Amherst %
\and Cindy Xiong\thanks{e-mail: cxiong@u.northwestern.edu}\\ %
     \scriptsize Northwestern University %
\and Steven Franconeri \thanks{e-mail: franconeri@northwestern.edu}\\ %
     \scriptsize Northwestern University %
\and Alberto Cairo\thanks{e-mail: a.cairo@miami.edu}\\ %
     \scriptsize University of Miami %
\and Narges Mahyar\thanks{e-mail: nmahyar@cs.umass.edu}\\ %
    \scriptsize UMass Amherst }

\teaser{
  \centering
  \includegraphics[width=0.9\linewidth,keepaspectratio]{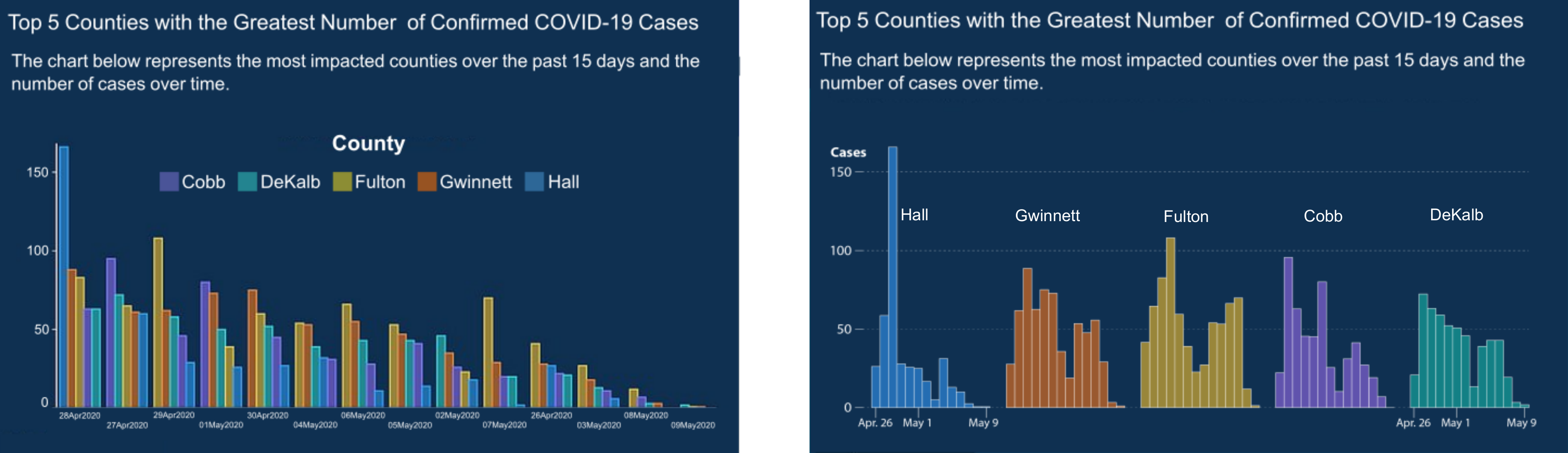}
  \caption{How do design decisions affect levels of understanding of the data in a visualization? We propose a systematic framework to evaluate data visualization across multiple facets of understanding based on Bloom's taxonomy.}
  \label{fig:teaser}
}

\abstract{ Understanding a visualization is a multi-level process. A reader must extract and extrapolate from numeric facts, understand how those facts apply to both the context of the data and other potential contexts, and draw or evaluate conclusions from the data. A well-designed visualization should support each of these levels of understanding. We diagnose levels of understanding of visualized data by adapting Bloom's taxonomy, a common framework from the education literature. We describe each level of the framework and provide examples for how it can be applied to evaluate the efficacy of data visualizations along six levels of knowledge acquisition - knowledge, comprehension, application, analysis, synthesis, and evaluation. We present three case studies showing that this framework expands on existing methods to comprehensively measure how a visualization design facilitates a viewer's understanding of visualizations. Although Bloom's original taxonomy suggests a strong hierarchical structure for some domains, we found few examples of dependent relationships between performance at different levels for our three case studies. If this level-independence holds across new tested visualizations, the taxonomy could serve to inspire more targeted evaluations of levels of understanding that are relevant to a communication goal. 

} 

\CCScatlist{
  \CCScatTwelve{Human-centered computing}{Visu\-al\-iza\-tion}{Visualization design and evaluation methods}{};
}

\begin{document}


\firstsection{Introduction}

\maketitle

When designers create visualizations for communication, they make choices about encoding and design that they think will accurately and persuasively communicate their interpretation of the data. The ultimate interpretation of a visualization depends on both the designer and the reader \cite{dork2013critical}. Because visualizations are being used to communicate data about real-world problems and phenomena, it is important for visualization designers and practitioners to validate if their design decisions are successful \cite{santos2008evaluating}.

Consider the two graphs shown above in Figure \ref{fig:teaser} which contain the same underlying data and differ only in the way the bars are organized. What do you notice when you look at each? How might we systematically test the impact of this reorganization on a typical viewer's understanding? We could ask how quickly and accurately a viewer extracts specific points, but we might expect to see no difference between the charts. In contrast, we would expect differences in how viewers describe their overall takeaway message (perhaps ``COVID-19 cases are decreasing'' for the left and ``COVID-19 cases are not decreasing everywhere'' for the right) or extrapolate predictions (``there will be no new COVID-19 cases on May 10'' for the left and ``there will be some new cases on May 10'' for the right). What other differences in understanding might you predict? 

We sought a more systematic way to evaluate these levels of understanding and how each level might differ depending on a visualization's design -- understanding is far more than extraction of individual values, after all. 
Policy makers rely on visualized data to decide which public project to fund. Health officials rely on visualized data to monitor and predict the trends of a pandemic. Investors rely on visualized data to evaluate the success of their businesses and identify future directions. 
The public sees visualizations in the news and must extrapolate how the data relate to their community.  In all of these cases, viewers must synthesize the visualized information and evaluate its soundness, then extrapolate the knowledge gained to other situations. An effective visualization should facilitate these interactions with data. 

Existing evaluation methods tend to focus on perceptual tasks such as retrieving values or comparing means \cite{lee2017vlat}. The visualization community has acknowledged that these tasks are effective for measuring perceptual accuracy, but insufficient for evaluating the knowledge or insight obtained \cite{north2006insight, kosara2016sand}. 
Visualization researchers have since introduced techniques to evaluate understanding beyond perception, such as answering factual questions about visualizations \cite{wang2019comics}, summarizing key messages \cite{bateman2010junk}, or  reporting what was learned \cite{walny2014sketches}.
Though these techniques can evaluate some aspects of the knowledge-generation process, the field still lacks a systematic method for evaluating the affordances for understanding that are provided by visualizations. 
We have seen in the visualization community that frameworks for evaluation (such as \cite{AATF}) can be helpful for systematizing the design of experiments, so we looked outside of the discipline for inspiration.

Inspired by evaluation techniques from the field of education \cite{mahyar2015towards}, we expand on the existing methods in the visualization community and introduce a comprehensive framework for evaluation based on Bloom's 6-level taxonomy.
Though the original taxonomy suggests a strict hierarchy between the levels, our results are consistent with challenges to a strict hierarchical structure (e.g. \cite{bagchi2014hierarchy}), finding some dependencies among levels, but also independence among others.

With our proposed framework, we systematically construct a set of questions, moving from testing a person's perceptual understanding of a visualization to eliciting how they would apply the learned information from the visualization to a real-world problem.
We demonstrate the applicability of our proposed framework with three case studies comparing two alternative designs of real-world visualizations, selected because they represent several common chart types and were identified by the data visualization or data journalism communities as being confusing or misleading.
We found that different versions of each chart afford different percepts and conclusions across particular levels of the taxonomy, allowing a systematic evaluation of similarities and differences between designs.
We also found interesting relationships between the levels of understanding measured, which suggest some level of dependence between skills.

The contributions of this work are the following: (1) a novel framework that provides a systematic way to evaluate levels of understanding in a visualization; (2) three ''in the wild'' case studies that demonstrate how this method might work; and (3) a discussion of future work and improvements in this area of research.

\section{Measuring Understanding in Visualization}

Within the visualization community, there are a variety of metrics used to assess particular aspects of understanding. 
Critically, this includes the common metrics of response speed and accuracy. 
While the relevance of the accuracy component is fairly straightforward, Chen et al \cite{Chen2014WhatFor} argued that if visualizations are intended to make content easier to understand, then ``saving time'' is a useful metric of measuring success. 
While this is true, we argue that visualizations are not just intended to help people understand information easier or faster, but also to better afford aspects of understanding such as applying or evaluating the visualized information.
Therefore, in order to truly evaluate the efficacy of a visualization, we need to move beyond purely quantitative measures and ask more difficult, open-ended questions \cite{north2006insight} which can capture different information than closed-ended methods \cite{north2011comparison, reja2003open}.

There are several existing methods which in some way accomplish this goal.
For example, one study asked participants to describe one thing they found interesting or surprising \cite{walny2014sketches}. 
Collecting salient points is a very direct way of assessing what participants found important, but only requiring one insight may mean that the information reported is not necessarily indicative of all that they have learned. 
Further, while it measures what the perceived takeaway is, it is not able to assess other related levels of learning, such as how the knowledge could be applied to a new situation; a step critical to the learning process \cite{perkins1992transfer}.

Another approach asked participants to describe everything that came to mind as they interacted with a visualization \cite{north2011comparison}.
Due to the open-ended nature of this method, it is certainly thorough, but it may not be possible in remote settings and relies on the ability of the participant to verbalize their thoughts reliably.
Additionally, one downside of the approach (as identified in the original paper) is the sheer volume of qualitative information that is generated \cite{north2011comparison}.
Our approach considerably reduces this burden, condensing the amount of information collected into only six questions.

A third approach in this vein asked participants to describe different aspects of a visualization such as the overall message, trend, and sentiment \cite{bateman2010junk}.
By requiring participants to comment on each aspect of interest, this method is effective at assessing the breadth of knowledge obtained by participants, but it too misses the opportunity to assess other kinds of knowledge acquisition which might have revealed other differences in the stimuli.

As we have summarized, while some researchers are asking more difficult questions that assess particular aspects of understanding, \textbf{they all lack a system for comprehensively evaluating different levels of understanding}. Such concrete models can also help to measure affordances that are provided by visualizations to help readers obtain knowledge. This will enable visualization designers to assess their readers beyond graph literacy and create more effective visualizations for their audience.
As visualization is still a relatively new field, we can take inspiration from the effective methods used in other disciplines.

\section{Measuring Understanding in Other Disciplines}
In other disciplines, there are a larger variety of methods used to assess understanding.
In visual literacy research, questionnaires comprised of multiple choice questions have been used to quantify how fluently people understand graphics in general (e.g. \cite{lee2017vlat}) and in medical contexts (e.g. \cite{galesic2011graph, okan2019using}).
We see evidence of similar use of questionnaires in Medicine and Chemistry to, for example, measure patients' understanding of Informed Consent documents \cite{joffe2001quic} and assess the public's understanding of nanotechnology \cite{schonborn2015NanoKI}.

Other than quantitative questionnaires, research has also used more qualitative or open-ended measures.
For example, benchmarks have been used to measure knowledge of topics important to the job performance of professional psychologists and to evaluate the way that this knowledge appears in on-the-job behavior \cite{fouad2009benchmarks}.
In another approach, researchers used a set of open-ended questions about Climate Change to establish the effect of combining refutation texts with graphics and analogies on understanding \cite{danielson2016refutation}.

Inspired by the work done in other disciplines and the need for a systematic way to measure knowledge-related affordances in visualization, we endeavored to create a method which combines open-ended questioning techniques, which have been shown to capture data that would otherwise have been lost \cite{reja2003open}, with a comprehensive survey of different aspects of understanding. 
Thankfully, education has provided us with a helpful framework to do just this -- Bloom's six-level taxonomy of educational objectives.
In 2015, Mahyar et al proposed the use of Bloom's taxonomy in information visualization to measure the depth of engagement and knowledge obtained by viewers \cite{mahyar2015towards}. However, as of yet, we do not have a concrete description of how it can be applied and a demonstration of its efficacy. In Section \ref{blooms}, we describe Bloom's taxonomy in detail and provide specific examples of how this framework can be used to assess visualizations.

\section{Existing Taxonomies in Visualization}
Existing taxonomies within the visualization community can be roughly divided into three categories: taxonomies of visualizations, objectives, and actions. 
Where taxonomies of visualizations classify types of data visualizations (e.g., \cite{buja1996interactive, pfitzner2003unified}), taxonomies of objectives categorize the questions that a user wants the answers to in order to solve a problem (e.g., \cite{brehmer2013multi, murray2017taxonomy, zhou1998visual, valiati2006taxonomy, chen2016optimize, rind2016taskCube, pretorius2014multi}). 
Finally, taxonomies of actions classify concrete actions done in pursuit of an objective (e.g., \cite{schulz2013design, ahn2014task, kerracher2015task, AATF, rind2016taskCube}).

Our proposed framework is a taxonomy of objectives and is most similar to the taxonomy of analytic tasks described by Amar and Stasko \cite{amar2004,amar2005}.
In both taxonomies, the categories of objectives correspond to types of tasks which need to be completed in decision-making practices. However, our proposed taxonomy is more comprehensive, expanding the scope of objectives to include both more simplistic tasks and the final conclusions drawn.

Some existing work also focuses on participant learning outcomes, but they cannot be used to form evaluation questions -- only to evaluate the responses. This work quantifies the completeness and correctness of a response (e.g., in \cite{bateman2010junk, xiong2019illusion}) or the complexity of a reported insight (e.g., in \cite{walny2014sketches}) in order to  evaluate open-ended responses in a post-hoc way.

\section{Bloom's Taxonomy}
\label{blooms}
In 1948, a group of psychologists and teachers had an informal meeting at the American Psychological Association Convention, brought together by a shared problem. They wanted their students to understand their lessons, but they each had different ideas about what they meant by terms like ``understand'' or ``internalize,'' leaving them with no way to compare those objectives \cite{bloom1956taxonomy}.
Motivated by a desire to find a common, rigorous way to categorize student understanding, the group came up with the Taxonomy of Educational Objectives, described in a handbook published in 1956. The system is now commonly referred to as Bloom's taxonomy, named after Benjamin Bloom, the educational psychologist who edited the handbook.

In the years since its creation, Bloom's taxonomy has been used broadly in Education \cite{Krathwohl2002Revision}.
Educators can create activities and assessments that target specific levels of the taxonomy \cite{arneson2018visual}
or use it to evaluate the assessments they already use to better understand the objectives being tested and potentially inspire a broadening of those objectives \cite{jones2009exams}.
In addition, Bloom's taxonomy has been applied widely in other fields such as Biology, Business, and Health Sciences, to create and evaluate exam questions \cite{crowe2008BBT}, teach critical thinking skills \cite{nentl2008business}, guide course creation on substance use disorders \cite{Muzyk2018substance}, and visualize the breadth of goals present while developing curriculum on visual literacy \cite{FaccinHermanVisualLit}.

Drawing inspiration from the taxonomies of Biology, Bloom's taxonomy is intended to be hierarchical, meaning that learning at the higher levels is dependent on demonstrating mastery of lower levels. 
However, this is not necessarily a realistic representation of the learning process, which is not linear \cite{bergerWrongBlooms}. Therefore, although we will now describe the levels of Bloom's taxonomy in the order originally presented, we reject the assumption that a strict hierarchy exists between levels.
Instead, we view them as complementary skills.
For a quick reference to the 6 levels and how they can be used for evaluating visualization, refer to Table \ref{table:blooms}.

\begin{table*}[hbtp]
\centering
\scriptsize
\setlength\tabcolsep{5pt}
\ra{1}
\begin{tabularx}{\linewidth}{l X X}
	\toprule
	\textbf{Level} & \textbf{Description} &\textbf{Example Tasks}\\ \hline
    Knowledge & Recall basic facts and definitions. &
    \begin{minipage}[t]{\linewidth}
    \begin{itemize}[nosep,after=\strut,leftmargin=*]
        \item Retrieve points
        \item Locate value
        \item Identify axis labels
    \end{itemize}
    \end{minipage}\\ \arrayrulecolor{lightgray}\hline
    Comprehension & Understand the information in context. &   \begin{minipage}[t]{\linewidth}
    \begin{itemize}[nosep,after=\strut,leftmargin=*]
        \item Summarize main message/take away
        \item Describe content of visualization
        \item Explain the topic of the visualization
    \end{itemize}
    \end{minipage}\\ \hline
    Application & Apply knowledge to a new problem or represent it differently. &  \begin{minipage}[t]{\linewidth}
    \begin{itemize}[nosep,after=\strut,leftmargin=*]
        \item Use a percentage and total population to calculate a number
        \item Calculate the difference between two points
        \item Translate the data in a chart to a table
    \end{itemize}
    \end{minipage}\\ \hline
    Analysis & Break down a concept into parts and understand their relationship. &  \begin{minipage}[t]{\linewidth}
    \begin{itemize}[nosep,after=\strut,leftmargin=*]
        \item Describe a trend
        \item Describe the relationship between two variables
        \item Identify what data was used to come to a conclusion
    \end{itemize}
    \end{minipage}\\ \hline
    Synthesis & Use knowledge to create something new. &  \begin{minipage}[t]{\linewidth}
    \begin{itemize}[nosep,after=\strut,leftmargin=*]
        \item Predict a future value
        \item Generate a new visual representation
    \end{itemize}
    \end{minipage}\\ \hline
    Evaluation & Judge the value of information, backed by evidence. &  \begin{minipage}[t]{\linewidth}
    \begin{itemize}[nosep,after=\strut,leftmargin=*]
        \item Justify a conclusion based on data
        \item Judge which design is more appropriate
    \end{itemize}
    \end{minipage}\\
	\arrayrulecolor{black}\bottomrule
\end{tabularx}
\vspace{2pt}
\caption{This table presents the 6 levels present in the original Bloom's taxonomy \cite{bloom1956taxonomy}, a short description of each, and example tasks specific to the visualization community.}
\label{table:blooms}
\vspace{-5mm}
\end{table*}

\subsection{Knowledge}
The Knowledge level historically describes the simplest learning objective demonstrable by the learner.
Associated with verbs such as retrieve, identify, and recall, at this level a learner is able to accurately \textbf{recall or recognize factual information} that they have learned.
Note that this does not require the learner to understand any contextual information or the reason behind facts \cite{bloom1956taxonomy}.

This level, in its original context, describes a very simple learning task -- reporting back something already seen. Therefore, for visualization, we translate this level into tasks which ask participants to locate and report specific pieces of information. In both the original and translation, no transformation is applied to the information by the viewer and no understanding of context is required.
In the experiment described later in this paper, we used this question to ask participants to locate specific data points, though other appropriate tasks might include copying text from an annotation layer or identifying what manipulations an interactive visualization offers.

\subsection{Comprehension}
At the Comprehension level, learners begin to \textbf{understand the underlying information as a whole} \cite{bloom1956taxonomy}.
Traditionally, questions at this level ask learners to write summaries or identify key ideas \cite{dalton1986BloomsExamples} and use verbs such as describe, explain, and summarize.

When applied to visualizations, we can translate this level's focus on understanding information as a whole into tasks that ask about features present in a dataset as a whole. For example, we suggest asking for a general summary of the data or the key take-away messages. Questions formed at this level can be more open-ended than those at the Knowledge level and, through this, can reveal the different conclusions afforded by the visualization being evaluated.

\subsection{Application}
At the Application level, learners \textbf{apply their knowledge} to solve an unfamiliar problem \cite{bloom1956taxonomy}.
This level is commonly associated with verbs such as translate, solve, calculate, and apply.

For visualization, this level could be translated to tasks where the participant solves a problem using the data from the visualization, such as identifying a proportion and using it in a simple computation. This approach may be most appropriate when the response modality for the participant is restricted to text. In our experiment, we asked participants to determine the difference between two data values. 
For situations where there is not an obvious problem to be solved, this level can also be interpreted as translating knowledge from one form to another.
For example, one could ask participants to translate the data displayed in a visualization into another visual style, though this approach requires a more complicated response modality.

\subsection{Analysis}
At the Analysis level, the learner is expected to \textbf{break down a topic into parts and understand the relationship between each part} \cite{bloom1956taxonomy}.
This level is therefore associated with verbs such as classify, break-down, associate, and relate.

Questions targeting this level could ask about trends, as this requires the participant to identify relevant components and then compare their spatial relationship to each other. We rely on this type of question in the experiment presented in this paper. 
Alternately, questions could also ask participants to identify which pieces of evidence were used to support a specific conclusion drawn from the data. 
This translation views the conclusion as the ``topic'' to be broken down and the data points as the components. Acquiring a conclusion from the data points requires understanding the relationship of the points to each other.

\subsection{Synthesis}
Where the focus of Analysis level was to test the learner's ability to decompose a topic into its requisite parts, the Synthesis level focuses on the learner's ability to \textbf{put ideas together to create something new} \cite{bloom1956taxonomy}.
Among the hierarchy, this is the first level which requires a certain amount of creativity from the learner and is associated with the verbs such as create, invent, predict, and devise \cite{bloom1956taxonomy, dalton1986BloomsExamples}.

When applied to visualizations, questions targeting this level could ask participants to make predictions about what values will come next in a sequence.
In this translation, the participant takes existing trends and values extrapolates on them to form a prediction.
Alternately, in the case of interactive visualizations, participants could instead use interactive features to find a view of the data which reveals something new. 
Because we used static data visualizations in our experiment, we used the prior approach and asked participants to make a prediction. 

\subsection{Evaluation}
Finally, we arrive at the sixth and final level of Bloom's taxonomy -- Evaluation.
This level evaluates a learner's ability to
\textbf{judge the value of a topic or idea} based on criteria that is either provided or self-derived \cite{bloom1956taxonomy}.
Therefore, rather than judgements, tasks at this level may look more like arguments or proofs, as evidenced by the verbs often associated with this level which include judge, justify, argue, and recommend \cite{dalton1986BloomsExamples}.

The most straightforward translation of this level to an evaluation task might be to ask participants to judge the quality of the visualization itself by some provided criteria (e.g., reliability). This method might be appropriate if the experiment aims to evaluate the participant's understanding of the visual encoding of the visualization.
Alternately, if the experiment aims to evaluate the participant's understanding of the underlying data, we suggest instead either asking participants to come to a conclusion and provide a data-based justification for that conclusion or to provide a conclusion and ask participants to only provide the justification.
With this tactic, participants judge the value of data features when deciding which are appropriate to justify the conclusion drawn.
We use this approach in our experiments.

\section{Experiment}
To demonstrate the kind of results an experiment could obtain with our evaluation method, we evaluated three pairs of visualizations as case studies.

\subsection{Stimuli}
For our stimuli, we selected 3 static, real world data visualizations varying in complexity and design that were identified by the internet community as being confusing or misleading \cite{GeorgiaCOVIDConfusing, CAImmigrationConfusing, econMistakes}. We also collected three corresponding redesigns of the confusing visualization created either by one of the authors who is an expert in visualization design or the original visualization designer at the Economist \cite{econMistakes}, where the goal of the redesign was to clarify the message conveyed by the visualization. 

\begin{figure}[hbtp]
    \centering
    \includegraphics[width=\linewidth,keepaspectratio]{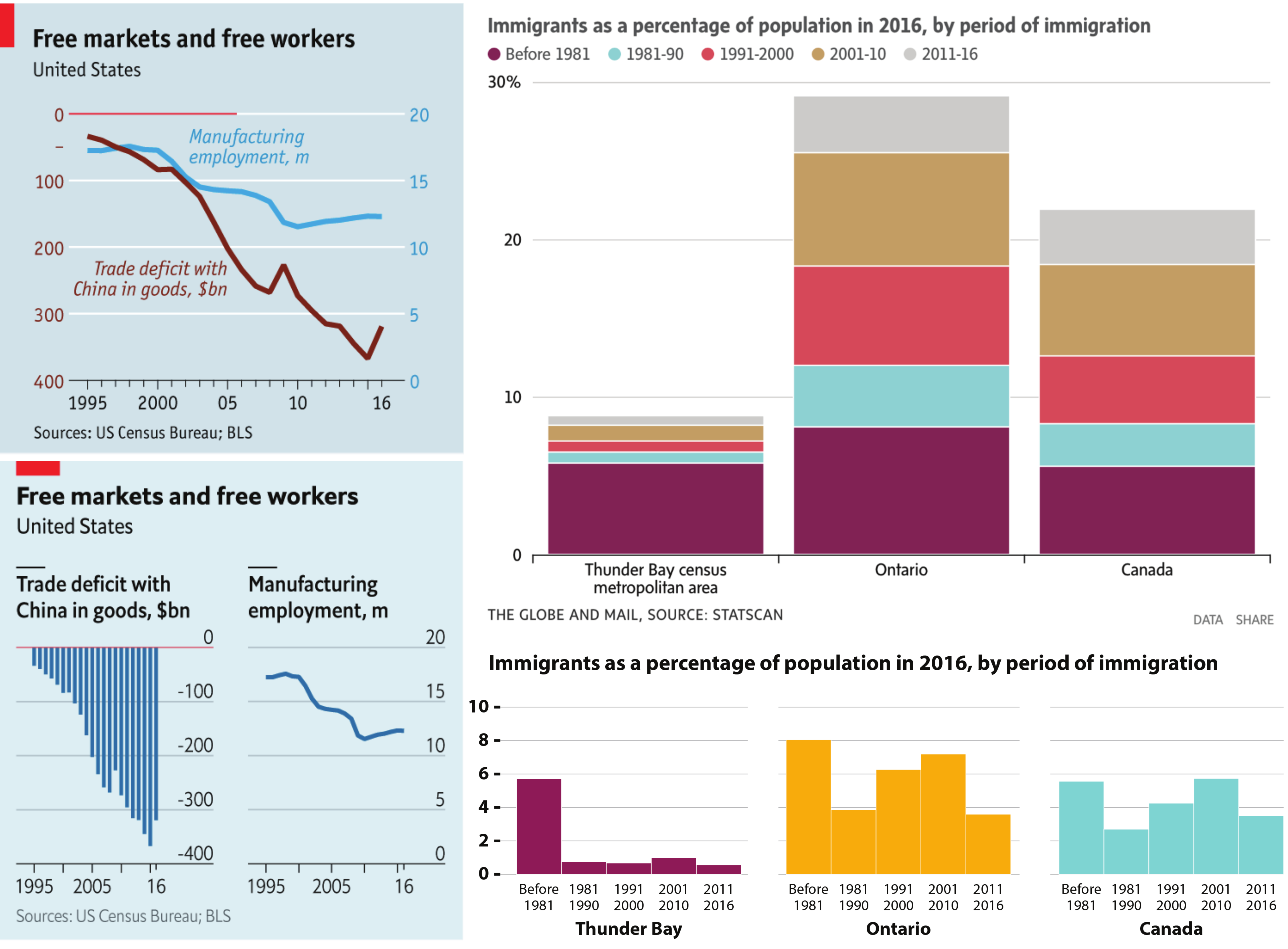}
    \caption{The Markets (left) and Immigration (right) charts used in our experiment. The original versions are on top.}
    \label{fig:charts}
\end{figure}

\vspace{4pt}\noindent\textbf{Markets}
The first stimulus (as shown on the left in Figure \ref{fig:charts}) was highlighted in an online article from The Economist titled ``Mistakes, we've drawn a few'' \cite{econMistakes}. The original chart depicts the US trade deficit with China and the number of people in the US employed in manufacturing between 1995 and 2016 as two line plots. The cardinal sin committed by this chart is its double y-axes -- one positive and one negative, color-coded to match its associated line. In the article, the author presents a redesigned version of the chart that separates the two plotted quantities into separate, side-by-side charts and encodes the trade deficit using bars connected to the 0 baseline to emphasize the directionality of the plot.

\vspace{4pt}\noindent\textbf{Immigration}
The second stimulus was produced by The Globe and Mail to discuss differences in immigrant populations in Thunder Bay, Ontario, and Canada (as shown on the right in Figure \ref{fig:charts}).
The original version of the chart features three stacked bars, one per location. 
Each stack is divided into five sections, corresponding to the decade when people immigrated to Canada.
At a first glance, this chart might be confusing because the sections correspond to time -- a variable conventionally reserved for the x-axis. 

Unlike the first chart, the structural problem with this chart is harder to identify. 
If the purpose of the chart is to highlight the differences in distribution or visualize the aspect of time, the original chart does a poor job.
However, if the primary purpose of this chart is to highlight differences between the total percentage of immigrants in each area, the original chart might accomplish that goal.
For the purpose of this paper, we assumed that the designer of this chart intended to show the differences between the distributions of immigrant year of arrival and emphasized this message in our redesigned chart by unstacking the bars and recoloring the chart such that each location uses a different color (instead of each decade).

\vspace{4pt}\noindent\textbf{COVID-19}
The final chart we selected was a now retracted chart created by the Georgia Health Department\cite{GA_DPH} depicting the number of COVID-19 cases reported over two weeks in the five counties with the largest number of cases (see Figure \ref{fig:COVID}).
The original version of this chart at first looks uncomplicated. It's just a simple bar chart, but a close reading of the labels along the x-axis reveals the problem -- the dates are not in chronological order.
Instead, both the days and the counties within each day are sorted by severity. 

As made obvious by its retraction, if this chart was intended to show how the number of cases of COVID-19 have changed over time, this chart is plainly misleading.
However, like the chart on Canadian Immigration, this chart would be appropriate for answering a different question.
Namely: which days saw the largest or smallest number of COVID-19 cases reported, and which counties had the highest or lowest number of cases within each of those days?
Unfortunately, this message is not supported by the annotation layer of the chart, which uses the phrase ``the number of cases over time."
For comparison, one of the authors created a second version of this chart with its dates in chronological order and with the bars divided by county to make it easier to compare how the number of cases changed over time within each region.

\subsection{Methods}
We conducted our experiment on Amazon's Mechanical Turk and recruited a total of 60 Workers ($\textrm{Mean}_\textrm{age} = 36.4$, $\textrm{SD}_\textrm{age} = 10.7$, 18 women, 41 men, 1 other).
In line with past research on acquiring quality results without attention check questions, we required that workers had completed at least 100 tasks with an approval rate of at least 95\% \cite{peer2014reputation}.
The experiment took approximately 30 minutes and participants were compensated with \$5.00 for their time.

In this experiment, participants were shown 3 charts and asked to answer 6 questions based on each one.
Each question was designed to target a specific level of Bloom's taxonomy and was presented in order, beginning with Knowledge and ending with Evaluation.
The order of charts was determined via a 2 by 3 Latin Square design (yielding 6 unique chart orderings and thus 6 conditions). 
Each participant saw 1 version of each chart in accordance with their assigned condition.
In 3 of the conditions, participants saw 2 of the original charts and 1 redesigned chart, while participants assigned to the other 3 conditions saw 1 original and 2 redesigned charts.

We want to emphasize that the purpose of using the alternative designs in this study was to show the range of reader interpretations and visualization affordances our method could evaluate, rather than to generate concrete design guidelines from these case studies. Because of this, we note that our analysis utilizes an exploratory approach -- not formal hypothesis testing.

\section{Results}
\begin{figure*}[ht]
    \centering
    \includegraphics[width=\linewidth]{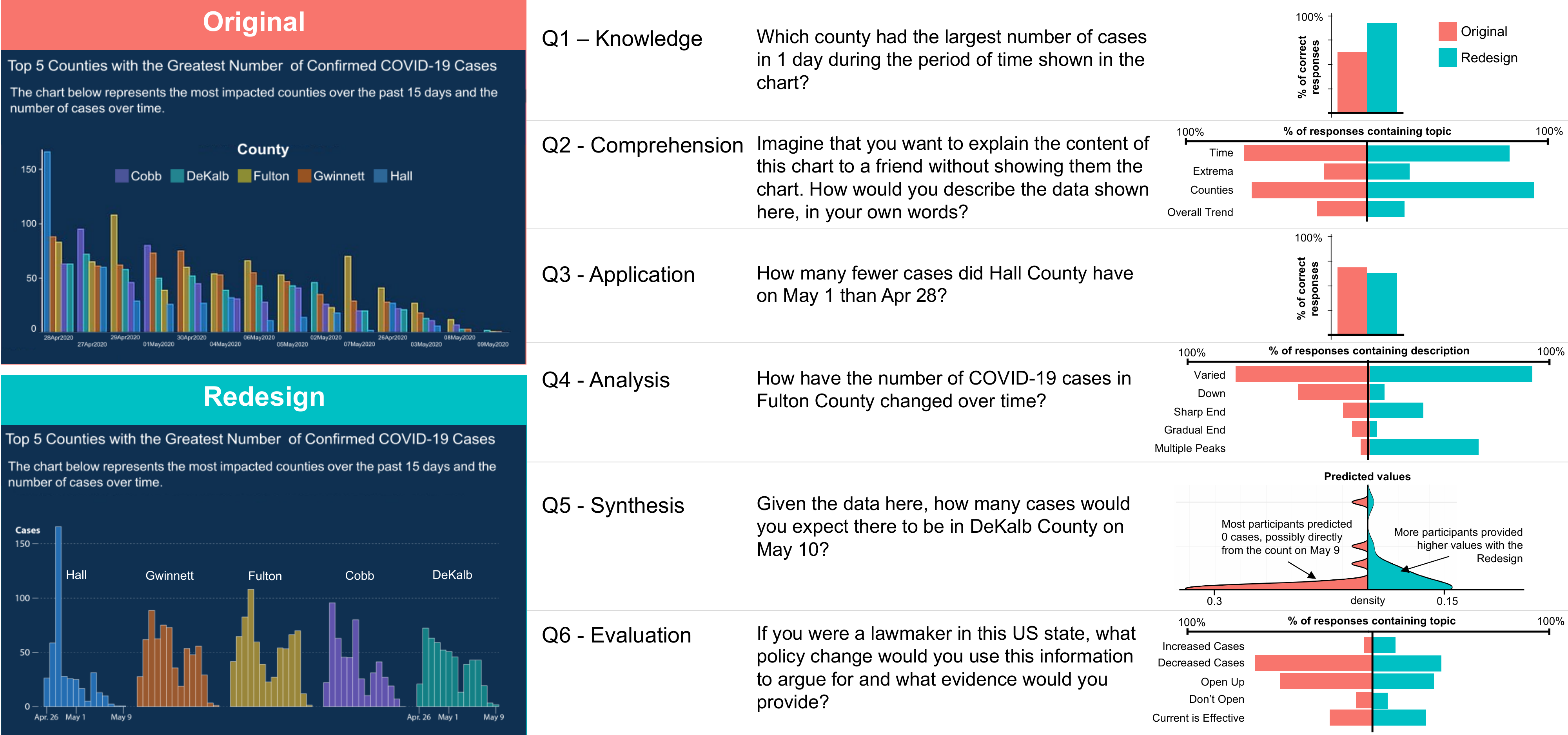}
    \caption{A summary of the stimuli and questions used in the experiment for the COVID charts. Charts on the right show results for each level examined. Participants answered Question 1 correctly significantly more often when using the redesigned chart than the original. This does not hold for Question 3, however. When describing the contents of the chart (Q2), participants viewing the redesigned chart were more likely to talk about the chart on a county level, but otherwise answered similarly. Participants more often incorrectly classified the trend of the chart as going down when viewing the original version and were less likely to comment on the bi-modal shape (Q4). We observed no effect of the version on the prediction made about the number of cases (Q5) and, unexpectedly, participants argued for ``opening up'' and cited decreasing cases as frequently in both groups (Q6).}
    \label{fig:COVID}
\end{figure*}

Figure \ref{fig:COVID} provides the questions used for the COVID-19 chart and a summary of results for all 6 levels. 

\subsection{Question 1: Knowledge}
We begin our analyses with the first question, which asked participants to locate particular values in the chart. We constructed a logistic regression predicting response accuracy with visualization design (original vs. redesigned) across the three chart topics. We found that participants were more likely to answer the question correctly when they saw the redesigned visualization compared to the original ($\chi^2(1) = 21.45$, $p < 0.001$). 
There was no main effect of the topic on this relationship ($\chi^2(2) = 4.54$, $p = 0.10$) (see Supplemental materials for pair-wise odds ratios).
This suggests that participants were retrieving values from the redesigned visualizations more accurately compared to the original version. 
Further, as the visualization was redesigned following common guidelines to increase clarity and afford more accurate value retrieval, this result suggests that the Knowledge level questions were successful in capturing the improvement brought by the redesign.

\subsection{Question 2: Comprehension}
\label{q2}
Questions targeting the Comprehension level were open-ended without defined correct or incorrect answers. Therefore, to analyze these responses, we read through the responses blind to the version seen, and identified varying categories of conclusions. Each response was tagged as either containing or not containing each conclusion. 
This approach was selected in order to compare differences in patterns identified between the two versions of each chart. We note that while this approach was chosen for demonstration, this kind of analysis is not specific to our proposed method; many existing techniques for analyzing qualitative results could be appropriate here, such as content-based analyses or interpretive analyses (see \cite{bazeley2013qualitativ} for some common practices). We conducted exploratory analysis using Pearson's chi-squared tests of independence to examine the relation between visualization design and user-identified salient patterns, with Yates' continuity correction and Bonferroni adjustments. See the Supplemental materials for details.

As shown in Figure \ref{fig:COVID}, participants who saw the original and redesigned versions of the COVID chart describe salient features with varying frequencies. For example, they were trendingly more likely to identify and compare the counties included in the redesigned chart, ($\chi^2(1,\, N = 54) = 5.33$, $p = 0.083$). 
The participants also mentioned different dates depending on the chart they viewed.
Namely, all seven participants who saw the original chart incorrectly identified the earliest date present as April 28, when it is April 26. This was a fact correctly identified by all participants who viewed the redesigned chart. 

In the Immigration chart, participants pointed out similar salient patterns in both visualizations. In other words, there is no difference in the distributions of conclusions drawn across both the original and the redesigned visualizations ($\chi^2(1,\, N = 51) = 0.51$, $p = 1$). 
We found this to be somewhat unexpected. We expected participants to more often comment on the total number of immigrants when the data was visualized as a stacked bar chart (as in the original) because they are viewed as the best choice for communicating an overall quantity among bar chart variations \cite{streit2014bar}, but this was not the case.
Similar to the immigration chart, we found that participants pointed out similar patterns in both versions of the Market chart (see Supplemental materials for more details).
 
These three case studies illustrate different outcomes to expect when analyzing responses to this kind of question. 
We see that the Market and Immigration redesigns did not differ from the original, suggesting that they afford the same set of conclusions.
This is different from the redesign of the COVID-19 visualization, which made a different set of patterns more salient to the viewers. 
We recognize that there could be more subtle differences in affordances between the original and the redesign that this question does not capture.
We speculate why this might be in the Discussion section.

\subsection{Question 3: Application}
Questions at this level challenged participants to determine the numerical difference between two specific points.
As with Question 1, we constructed a logistic regression predicting response accuracy with the original and redesigned visualization across the three charts.
Unlike with Question 1, participants who viewed the redesigned version of a chart were not statistically more likely to answer this question correctly ($\chi^2(1) = 2.51$, $p = 0.11$).
In addition, we observed no statistical difference between the accuracy between charts ($\chi^2(2) = 2.96$, $p = 0.23$) and no significant interaction ($\chi^2(2) = 2.97$, $p = 0.23$).

Although our questions at this level failed to find a significant difference between the two versions of the charts with respect to answer correctness, there may be other factors at play as well, which we discuss in the Discussion section.

\subsection{Question 4: Analysis}
\label{q4}
As with Question 2, questions for level 4 were open-ended. 
Participants were asked to describe a specific trend present in each chart. We identified a series of descriptions for each chart that either were mentioned by the participants or were determined by the authors as reasonable conclusions to mention. Their responses ranged from describing the directions of the trend (e.g., up or down) to commenting on modality.
We coded each response blind to the chart version, tagging each description as present or not (e.g., was the trend described as positive or not). We then compared the frequency of descriptions between the two chart versions.
We note that, as before, this is not the only way to complete this analysis, but was selected for demonstrative purposes.
We conducted chi-square analysis to examine the relation between chart design and chart descriptions, similar to that in Section \ref{q2}.

In response to the COVID-19 chart, we can see two distinct points of difference in the distributions (as shown in Figure \ref{fig:COVID}).
First, although participants were equally likely to describe the trend direction as containing both positive and negative sections,
($\chi^2(1,\, N=43) = 1.06$, $p = 1$), more participants incorrectly categorized the trend as decreasing when they viewed the original chart compared to the redesigned chart, although not significantly ($\chi^2(1,\, N=43) = 4.25$, $p = 0.20$). 
In addition, participants more frequently noticed the bi-modal shape in the data when viewing the redesigned chart ($\chi^2(1,\, N=43) = 12.34$, $p = 0.0022$), such that only one participant who viewed the original version identified the bi-modal shape in the data whereas over 60\% of the participants with the redesigned chart noticed this bi-modal pattern (see Figure \ref{fig:COVID}). This suggests that the redesigned version more readily affords viewers the ability to see the bi-modality of the case count.
In reality, answering this question correctly with the original version of the COVID-19 chart is extremely difficult. Because the labels on the x-axis are not in chronological order, characterizing the distribution over time would require mentally reorganizing the bars. However, it is important to note that none of the participants who saw the original version of this chart gave any indication on any question that they noticed that the dates were out of order, suggesting that either no participants noticed this feature or found it pertinent.

With the Immigration chart, we observed that most of the topics commented upon by participants are similar across versions, but with one very distinct difference.
Namely, participants who viewed the original chart were significantly more likely than those who viewed the redesign to \textbf{correctly} describe the trend of immigration as increasing ($\chi^2(1,\, N=51) = 10.23$, $p = 0.007$).
We suspect this unexpected disparity may be driven by the increasing size of the stacked bars over time.
Additionally, because of the relatively small change each decade, it is possible that participants viewing the redesigned chart considered the change not remarkable enough to mention. 
Regardless of the cause, it represents a distinct difference in the affordances between these charts that this question was able to capture. The distribution of topics discussed in response to the Markets chart was highly similar across visualization versions with no significant difference with respect to the version viewed (see Supplemental materials for details). 

The difference in distributions here suggests that the different versions of the chart lead to participants to come to describe the underlying data differently. This difference was not captured by the previous three levels, supporting the importance of a multi-leveled approach to evaluate visualization understanding.

\subsection{Question 5: Synthesis}
At this level, participants were asked to make predictions about future data values and trends for each chart. By looking at the distributions of the predicted values, we begin to unpack the decision making process afforded by each chart and version. In particular, we can glimpse both where the design had an effect on the average prediction made, as well as on the variance of those predictions. 
To identify statistical differences, we utilized Welch's two-sided t-tests with Bonferroni corrections to compare mean predictions and two-sided F-tests with Bonferroni corrections to compare variances.
For this analysis, we only included responses that contained a single, numeric answer, excluding those with ranges or that described a trend. This excluded 17\% of responses from the COVID-19 chart, 0\% from Immigration, and 7\% from the Markets charts.

For the COVID chart, participants were asked to predict the number of COVID-19 cases for one day beyond the dates shown in the chart. Results showed no significant difference between the mean prediction ($(M = [5.13,\, 5.13]$, $SD = [14.84,\, 11.77], \; t(41.83) = 0$, $p = 1$) nor the variance of the predictions made ($F(22, 22) = 1.59$, $p = 0.28$). This suggests that though we observed differences between the versions in previous questions, both charts afforded similar predictions to participants. 

For the immigration chart, participants were asked to predict the percentage of the population made up of immigrants in Thunder Bay, Ontario, and Canada. First, we observed that the mean prediction for all 3 locations were not significantly different across the two versions (see Supplemental materials for details).
While the variance of predictions between chart versions was not significantly different for Thunder Bay ($F(9, 7) = 0.73$, $p = 1$) or Ontario ($F(9, 7) = 4.75$, $p = 0.31$), it was significantly different for the predictions about Canada ($F(9,7) = 15.40$, $p = 0.009$), such that the predictions made about Canada's population were more clustered with the redesigned version than the original version. 
This suggests that while both chart versions afford similar numeric predictions, the redesigned chart affords less variation in predictions.

When viewing the Market chart, participants were asked to predict the trade deficit and the manufacturing employment.
Participants predicted trade deficits to be significantly higher when they viewed the redesigned version ($M = [272.19,\, 355.00], \; SD=[100.47,\, 49.61]$), $t(21.053) = -3.0032$, $p = 0.027$).
We additionally see that the shapes of the distributions were different: the predictions made with the original chart were highly clustered while those made with the redesigned chart were far more variable.
Our analysis suggests that the difference between these distributions is significant ($F(15,18) = 4.10$, $p = 0.021$).
There was no significant difference between the predictions made by participants about the number of people employed in manufacturing with respect to chart version ($M = [38.18,\, 11.56], \; SD = [57.60,\, 1.59]$, $t(16.026) = 1.9044$, $p = 0.30$).
However, we observed that the variance of the predictions is, again, significantly different between the two versions ($F(16, 15) = 1311.7$, $p < 0.001$).
These results suggests that while both charts afford similar predictions about manufacturing employment, the original chart affords significantly higher predictions about trade deficit than the redesigned chart. 

Although all of our questions about the Synthesis level did not reveal differences between versions of every chart, we were still able to reveal some different affordances that were not captured by the previous questions.
Evaluating these charts systematically across levels of understanding allows the identification of distinctions between affordances that might otherwise be missed.

\subsection{Question 6: Evaluation}
In the sixth and final question, participants were asked to apply their learning to a real-world situation by describing the argument that they would make.
Similar to Section \mbox{\ref{q2}} and \mbox{\ref{q4}}, we identified common conclusions and evidence that was cited, tagged each response as containing or not containing mentions of each of these topics, and used the same chi-square analysis approach.

In response to COVID-19 charts, the most common conclusion drawn by participants was for ``opening up'' (see Figure \ref{fig:COVID}).
This conclusion was mentioned by participants regardless of which version of the COVID-19 chart they were presented with ($\chi^2(1,\,N=44) = 0.76$, $p = 1$). 
Additionally, the most common evidence cited by participants in defense of their claim was a decrease in COVID-19 cases.
Surprisingly, participants were as likely to use this argument when viewing the original as the redesigned chart ($\chi^2(1,\, N=44) = 2.32$, $p = 0.76$).
This is unexpected, as the original COVID-19 chart shows a strong downward trend in cases (as long as the x-axis ordering is ignored), but the redesigned version of the chart shows increasing numbers in several of the counties.
This result suggests that despite the differences in design, both charts afforded participants the ability to come to the same conclusion.
The unexpected nature of this result further emphasizes the reason why asking difficult questions in evaluation is important; sometimes conclusions arise because (or in spite) of otherwise careful encoding.

For the Immigration chart, participants were asked to argue why the population of Thunder Bay was not representative of the population of Canada at large. The two main arguments were ``there were fewer immigrants in total" and ``there were fewer immigrants over time," but participants were equally likely to report these justifications regardless of the chart they saw ($\chi^2(1,\, N=47) = 0$, $p = 1$).

As for the Markets chart, we see that participants were very consistent across versions with respect to the conclusions drawn and evidence provided. 
One repeated theme present in the responses was a suggestion of a causal relationship between the number of people employed in manufacturing and the trade deficit with China.
This relationship was suggested in both directions (i.e. changes in the deficit causes changes in employment and changes in employment cause changes in the deficit) and did not appear statistically more often in response to either chart version ($\chi^2(1,\, N=46) = 0.014$, $p = 1$; $\chi^2(1,\, N=46) = 0.51$, $p = 1$).
This suggests that the two charts did not afford different conclusions nor justifications.

\subsection{Is this taxonomy really hierarchical?}
The original Bloom's taxonomy argues that there is a strict  hierarchical relationship between the levels \cite{bloom1956taxonomy}.
To shed some light on this idea, we looked for evidence that would suggest that performance on earlier questions is correlated with performance on later ones. 

Because Questions 1 and 3 had correct answers, we first examined if success on Question 1 was correlated with success on Question 3. We constructed a logistic regression and observed that participants who answered Question 1 correctly were significantly more likely to answer Question 3 correctly compared to those who answered Question 1 wrong ($\chi^2(1) = 22.3411$, $p < 0.001$) irrespective of chart topic ($\chi^2(2) = 1.02$, $p = 0.60$), version ($\chi^2(1) = 0.0099$, $p = 0.92$), and with no significant interaction effect ($\chi^2(2) = 3.46$, $p = 0.18$).
The effect size is quite pronounced for all three charts on both original and redesigned versions.
For example, participants who answered Question 1 correctly using the redesigned COVID-19 chart were 1.8 times more likely to answer Question 3 correctly.
More dramatically, participants who viewed the original COVID-19 chart were over 10 times more likely to answer Question 3 correctly if they answered Question 1 correctly.
This is unsurprising -- in order to answer Question 1 correctly, one must estimate one value and Question 3 builds on this skill by requiring participants to estimate two values and then subtract them. 

To further explore the relationship between performance on questions, we asked whether performance on Questions 1 or 3 predicted a participant's performance on Question 4. That is, is there a relationship between how well a person locates or determines the difference between points and how well they classify a trends?
For this, we reviewed the responses to Question 4 and marked those which were obviously incorrect.
We then used another logistic regression that predicted whether a response was obviously incorrect or not based on the chart topic, version, and the participants' performance on Question 1 and 3.
Our model suggests that there is no significant relationship between performance on Question 1 or Question 3 on the whether their response to Question 4 was reasonable or not (see Supplemental materials).
This suggests that though there may be some overlapping skills required between levels, this taxonomy is not hierarchical in nature.


\section{Discussion}

The visualization community needs better ways to evaluate what visualizations afford with respect to the understanding obtained by viewers. Being able to identify where affordances differ is critical to making smart design choices that enable different aspects of the knowledge acquisition process to occur. As we have demonstrated in this paper, the six levels of understanding from Bloom's taxonomy can provide a useful framework for generating new questions which comprehensively evaluate the affordances of visualizations across a spectrum of tasks and reveal differences in knowledge-making affordances that might otherwise have been missed.

Although we did not find differences between versions for every chart topic on every level, our evaluation method was able to capture some affordance differences, confirming the effectiveness of commonly-held beliefs about design and revealing under-explored directions in affordance evaluation. This method can complement perceptual tasks involving speed and accuracy by measuring what information readers can extract from visualizations and provides a systematic framework for designers to assess various levels of a reader's understanding. 


\subsection*{Limitations and Future Work}
We did not find significant differences between all design pairs for every level of understanding. There are several factors which might explain why we observed null effects in our case studies. First, the alternative visualizations were designed by domain experts, which may hold a different perspective from the average crowd-source worker in interpreting visualization \cite{xiong2019curse}. The reason we observed no differences in worker responses could have been because there exists no significant difference in affordances between the designs in the eyes of an average worker, despite what the experts thought. While it is likely that prior knowledge contributes to how well one performs on our task, we emphasize that we chose these alternative designs for the study to demonstrate the range of reader interpretations and visualization affordances that this method could evaluate, rather than to generate concrete design guidelines from these case studies. 

Additionally, we recognize that participants' ability to state a correct answer on the Knowledge, Application, and the Analysis levels may depend on their numeracy skills. We recommend that future researchers who use this method to evaluate visualization affordances also include participant numeracy evaluation in their experimental design (such as the one proposed in \cite{weller2013development}). Even without numeracy evaluation, we maintain that it is important to evaluate visualizations at the Knowledge, Application, and Analysis level to test whether the reader has correctly interpreted the visualization at a grammatical level. Further, the Application question is important because it is a measure of learning transfer which is an important but challenging aspect of the learning process. People often fail to transfer learning to a novel context \cite{perkins1992transfer}, which could explain why we did not observe differences in accuracy between chart versions in the application level -- this transfer task was difficult enough to overpower any effect of visualization design.
Finally, we recognize that the Evaluation level does not capture individual differences of biases in beliefs that may drive differing responses.



While we maintain that it is a worthwhile effort to apply Bloom's taxonomy to user-study task generation, we recognize that it does not always translate perfectly. First, Bloom's taxonomy was intended to evaluate learning that takes place over a much longer span than a typical person spends looking at a visualization. However, the levels in this taxonomy also relate to processes inherent in decision making procedures which also might occur over such short periods of time. 
Additionally, while we tried to capture the spirit of the levels of the original taxonomy, our translations may not measure identical skills. Some of the levels (particularly the upper-levels) seem difficult to evaluate in a purely textual format, but may be more easily translated to creation or editing tasks. 

Future iteration of this line of work could extend the list of example tasks in this paradigm to evoke specific, detailed responses that help designers and researchers gain insights regarding how a visualization reader is reacting to a visualization. We see strong potential for this method to evolve into a useful technique supplementing in-person interviews in remote environments where in-depth interviews may be difficult or impossible. Future researchers could also combine this evaluation method with other measures of graphical, linguistic, or numerical literacy to generate a more comprehensive evaluation method, or use this method to identify concrete design guidelines. Alternatively, researchers could diversify the data analysis approaches to extend our method beyond just evaluating affordances to cover graphical literacy or numeracy. For example, one could compare the participants' trend predictions in the synthesis (prediction) task to a ground truth to determine the accuracy of their prediction, which could inform researchers about their numeracy skills (e.g., how well the participant could interpolate/extrapolate trends in data). Additionally, although in our paradigm, we only compared two alternative designs of the same chart, this framework is flexible enough to allow for single or multi-chart comparisons.

\section{Conclusion}
Motivated by a desire to design data visualizations that communicate information accurately and effectively, the visualization community has long asked for novel ways to measure what is understood by the reader, and through this, what is afforded by visualizations. In this paper, we proposed a concrete framework grounded in Bloom's taxonomy and demonstrated how it can be used to form a set of questions that systematically evaluate the kinds of affordances provided by visualizations. We demonstrated how the framework can be used through 3 case studies of real-world visualizations and showed that our comprehensive method was able to identify understanding-related affordances that would have been missed by existing methods of evaluation that focuses on accuracy and speed. While it may not be appropriate to apply every level covered in the taxonomy to evaluate every visualization, this framework can help the community design questions that target specific aspects of the knowledge acquisition process that were previously unexplored or challenge existing assumptions about what makes one design choice ``better'' or ``worse'' than another. Finally, it allows us, for the first time, to systematically evaluate affordances in a way that is consistent with educational theories of learning and comparable across studies.


\bibliographystyle{abbrv-doi-hyperref-narrow}

\balance

\bibliography{template}
\end{document}